\documentclass[12pt,preprint]{aastex}

\begin{document}
\title{The X-ray Ridge Surrounding Sgr~A* at the Galactic Center}

\author{Gabriel Rockefeller\altaffilmark{1,2}, Christopher L.
  Fryer\altaffilmark{1,2}, Frederick K. Baganoff\altaffilmark{3}, and
  Fulvio Melia\altaffilmark{1,4,5}}

\altaffiltext{1}{Department of Physics, The University of Arizona,
  Tucson, AZ 85721}
\altaffiltext{2}{Theoretical Division, LANL, Los Alamos, NM 87545}
\altaffiltext{3}{Center for Space Research, Massachusetts Institute of
  Technology, Cambridge, MA 02139}
\altaffiltext{4}{Steward Observatory, The University of Arizona,
  Tucson, AZ 85721}
\altaffiltext{5}{Sir Thomas Lyle Fellow and Miegunyah Fellow}

\begin{abstract}
  We present the first detailed simulation of the interaction between
  the supernova explosion that produced Sgr~A East and the wind-swept
  inner $\sim 2$-pc region at the Galactic center.  The passage of the
  supernova ejecta through this medium produces an X-ray ridge $\sim
  9\arcsec$ to $15\arcsec$ to the NE of the supermassive black hole
  Sagittarius A* (Sgr~A*).  We show that the morphology and X-ray
  intensity of this feature match very well with recently obtained
  {\it Chandra} images, and we infer a supernova remnant age of less
  than $2,000$ years.  This young age---a factor 3--4 lower than
  previous estimates---arises from our inclusion of stellar wind
  effects in the initial (pre-explosion) conditions in the medium.
  The supernova does not clear out the central $\sim 0.2$-pc region
  around Sgr~A* and does not significantly alter the accretion rate
  onto the central black hole upon passage through the Galactic
  center.
\end{abstract}


\section{Introduction}

While Sagittarius A* (Sgr~A*) dominates the gravitational dynamics in
the central parsecs of the Galaxy, many other components are required
to explain the wealth of detailed observations of this busy region.
Even as we focus on the inner 3 parsecs, it is clear that other
classes of object contribute to the overall radiative emission from
the Galactic center (for a recent review, see \citealp{MF01}; see also
\citealp{Ma02}).  For example, the medium within which Sgr~A* is
embedded---bounded by the circumnuclear disk (CND, with inner radius
$\sim 2$--$3$~parsecs)---has a temperature $\sim 1.3$~keV and emits a
{\it Chandra}-detectable glow of diffuse X-rays \citep{Bag03}.
\citet{Roc04} have shown that this emission may be understood as the
result of mutual interactions between the winds of Wolf-Rayet and O/B
stars within $\sim 1$~parsec of the supermassive black hole.

The CND is perhaps the shredded remains of a giant molecular cloud
that passed by Sgr~A*.  The cavity within the inner $\sim
2$--$3$~parsecs of this structure may itself have been created by the
ablative influence of the cumulative wind outflow, which has by now
produced a bubble of hot, ionized plasma.  To make matters even more
complex, observations of the supernova (SN) remnant Sgr~A East suggest
that its broadband emission arises from shock heating in a recent
($<10,000$-year-old) supernova \citep[or other explosive outburst;
see][]{Me89} originating within $\sim 3$~parsecs of the black hole.

Any comprehensive model of this region must therefore include the
effects of all these components: the supermassive black hole, the
Wolf-Rayet and O/B winds, the dense CND and the expanding supernova
remnant, which we now see predominantly as a radio-emitting shell.
Although interacting winds can explain the bulk of the diffuse X-ray
flux from the inner $\sim 2$--$3$~parsecs, several other features seen
in the {\it Chandra} X-ray image are not as easy to explain without
the influence of some other interaction. In particular, a well-defined
ridge of X-ray emission just outside the central region, to the NE of
Sgr~A* (see Figure~\ref{fig:chandra}), may be evidence of an ongoing
collision between the SN ejecta and the cumulative Wolf-Rayet and O/B
winds emanating from within the cavity \citep{Ma02}.

In this {\it Letter}, we model the expansion of the supernova,
focusing on the effect this explosion has on the central few parsecs
surrounding Sgr~A*.  We then directly compare the X-ray emission
arising from the interaction zone with the actual {\it Chandra} image.

Interestingly, using the well-studied wind conditions at the Galactic
center, we may also be able to place tighter constraints on the
supernova explosion itself---both the released energy and the age of
the remnant.  With this knowledge, we can address several outstanding
issues pertaining to the influence of this explosion on the morphology
of the Galactic center: Did the supernova shock clear out the region
surrounding the black hole, effectively shutting down what would
otherwise have been a high accretion rate onto the black hole?  Could
the supernova have caused a brief increase in the accretion rate onto
Sgr~A*, producing a spike in X-ray emissivity that irradiated the
X-ray-fluorescing Sgr~B2 and other nearby molecular clouds some 300
years ago \citep[see, e.g.,][]{Sun98,From01}?

\section{General Physical Principles}

Our simulation uses the SNSPH smoothed particle hydrodynamics code
\citep*{FRW05} to follow the supernova explosion as it crosses the
Galactic center.  The domain of solution is a cube, $6$ pc on a side,
centered on the black hole.  Particles that move beyond this domain,
or within $1.9\times10^{17}$~cm of the origin (effectively onto the
black hole), are removed to simulate outflow (or inflow) conditions.
Our initial (pre-explosion) conditions are taken from the structure of
the wind-swept medium at the end of the simulation by \citet{Roc04}.
The initial particle distribution is constructed from a $\sim 1$
million particle supernova explosion placed 2~pc due east (in right
ascension) of the central supermassive black hole within a $\sim 6$
million particle wind-filled Galactic center region.

We assume that the density structure within this domain of solution at
the time of the supernova explosion is dominated by matter lost by the
Wolf-Rayet and O/B stars, plus the dense CND surrounding the central
black hole.  The CND is mimicked by 200 spherical clumps (totaling
$10^4\,M_\odot$), in a torus with a low filling factor surrounding the
black hole.  The winds from these stars (which we assume have not
changed noticeably in the past $10,000$ years) have blown a bubble in
the Galactic center that is probably at the edge of the 50~km~s$^{-1}$
molecular cloud \citep{Me89}.

Note, however, that our initial conditions do not include the initial
molecular cloud blown out by the stellar winds.  There is evidence
that the supernova shock has reached the boundary between the
wind-blown bubble and this cloud \citep{Yusef99}.  We can not address
these effects with this current set of simulations.  We also do not
include the effect of mass loss from the supernova progenitor itself.
However, the density structure near the X-ray ridge and the central
black hole is dominated by the $\sim 25$ stars we do include
\citep{Roc04}, and not by any outer stars or the surrounding molecular
cloud.

The structure of the supernova ejecta is set using a spherically
symmetric model from \citet*{Hun04}.  This $15\,M_\odot$ star is
exploded with an energy of $1.5\times10^{51}$\,erg; these properties
are typical of a supernova explosion, both in composition and energy.
We place the explosion into our domain of solution after the shock has
moved out for 1 year, at which point the explosion material is still
within $0.02$~pc of the supernova site.

\section{Calculations and Results}

\begin{figure}
\plotone{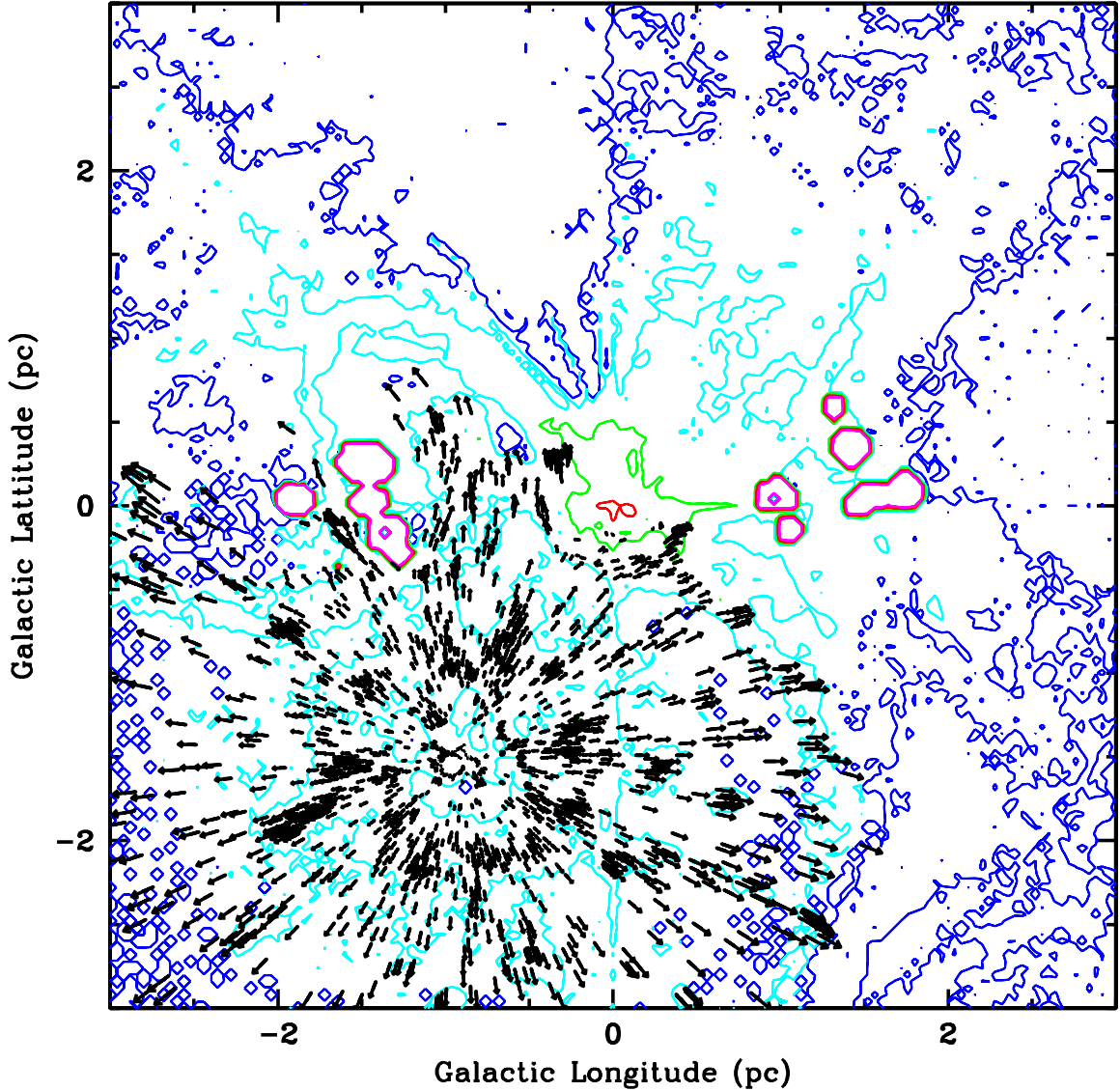}
\caption{Density contours of a 2-dimensional slice of the inner 3~pc
  surrounding Sgr~A* 650~y after the launch of the supernova
  explosion.  The contours take mean densities from a 0.2~pc slice
  centered on the radial distance of the supermassive black hole and
  correspond to number densities of 1 (blue), 10 (cyan), 100 (green),
  1000 (red), and $10^4$ (magenta) cm$^{-3}$.  The supermassive black
  hole is located at $0,0$.  Roughly at longitude = $-0.89$~pc,
  latitude=$-1.47$~pc is the origin of the supernova (at the same
  radial distance as the supermassive black hole).  The circumnuclear
  disk is modeled with several dense spherical clumps with a low
  volume filling factor (magenta contours).  The vectors show tracer
  particles of the supernova ejecta themselves (arrows denote velocity
  direction and length denotes velocity magnitude).  Note that the
  explosion flows around Sgr~A* and the dense central region along
  paths where the density is lowest.}
\label{fig:2ddens}
\end{figure}

The wind-swept initial conditions of the Galactic center lead to an
aspherical progression of the supernova shock.  Plowing through
diffuse regions with density $\sim 1$~particle~cm$^{-3}$, across the
dense CND, and through central regions with densities
$>10^4$~particle~cm$^{-3}$, the supernova shock is far from symmetric.
The ejecta flow around the dense regions, taking the path of least
resistance and producing shocks where they collide with and are
decelerated by the dense material.  Figure~\ref{fig:2ddens} shows the
density profile and position of a set of tracer supernova particles
650 years after the launch of the supernova explosion.  This is
roughly the time of deepest penetration of the supernova ejecta into
the region surrounding the supermassive black hole.  The actual
supernova shock has now moved beyond the Galactic Center and has
already passed beyond the southern and eastern edges of our simulation
grid.  The ejecta do not sweep through the central region, nor do they
significantly alter the density of the inner 0.2\,pc surrounding the
black hole.  A more detailed discussion of the effects of the
supernova shock on the central region, the implications for the black
hole accretion rate, and the possible excitation of Sgr~B2, will be
presented in \citet{Fry05}.

\begin{figure}
\plotone{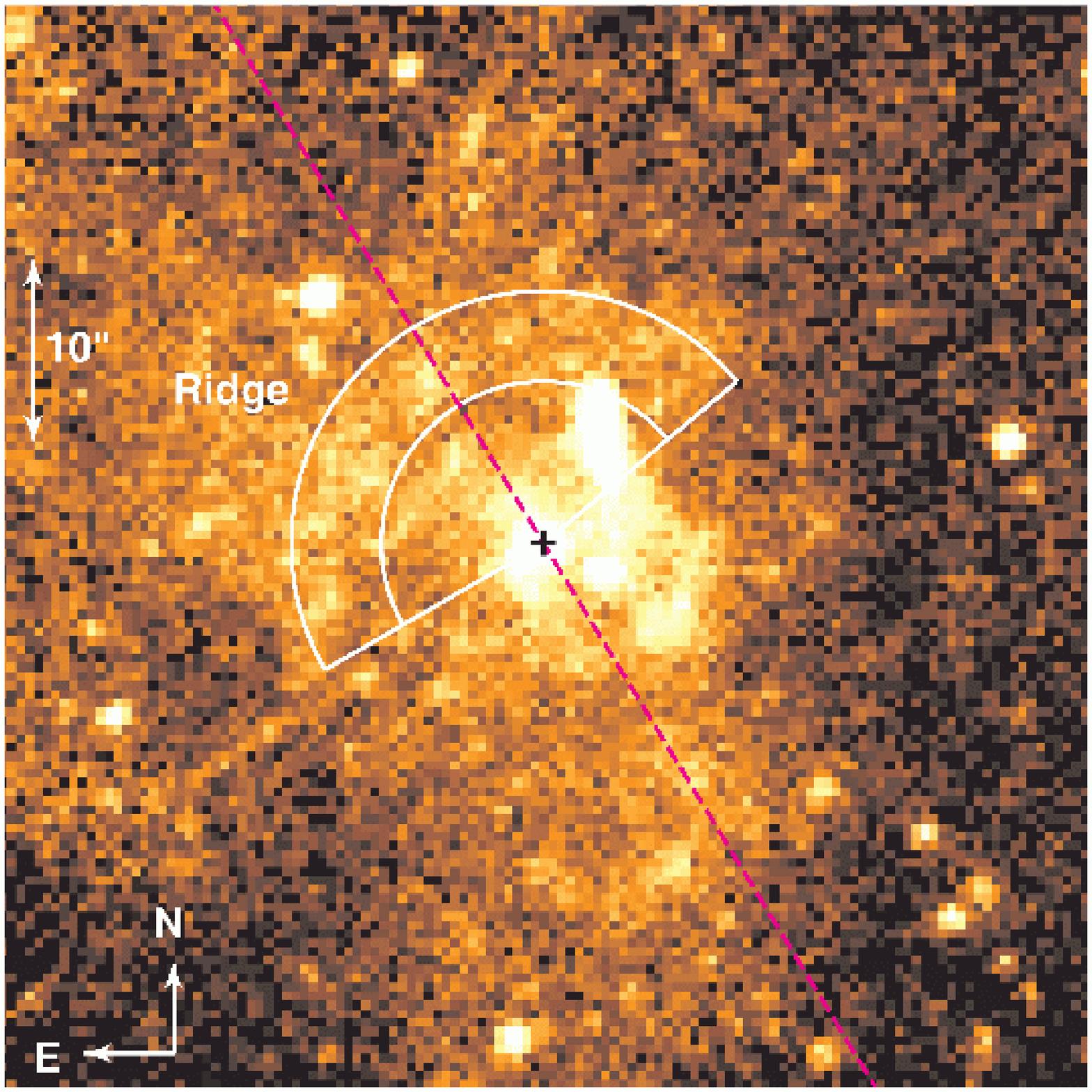}
\caption{{\it Chandra} image of the inner $1^{\prime}\times 1^{\prime}$ of 
  the Galaxy, centered on the location of the radio source Sgr~A*.
  The (square) pixel size is $0.\arcsec 492\times 0.\arcsec 492$, and
  the two arcs indicate distances of $9\arcsec$ and $15\arcsec$ from
  the center. The X-ray ridge, to the NE of Sgr~A* and in the
  direction of Sgr~A East, is bounded by these two arcs.}
\label{fig:chandra}
\end{figure}

\citet{Me89} estimated the age and energy of the supernova remnant by
assuming the supernova was plowing through a $10^4$~particle~cm$^{-3}$
molecular cloud.  They calculated a supernova remnant age of $7,500$
years and, to fit the observed shock temperature, they required an
explosion energy in excess of $10^{52}$\,erg.  However, using the
lower mean density in our wind-swept initial conditions allows us to
account for the observed remnant characteristics with a more typical
supernova explosion ($\sim 10^{51}$\,erg).  Our simulation also
suggests that the supernova remnant is younger than the estimate of
\citet{Me89} \citep[see][for more details on the fitting]{Fry05}.
From the position of the shock, we can set a crude lower limit to the
remnant's age at $\sim 1,000$ years, but a more precise answer may be
determined by comparing our predicted X-ray ridge properties to the
observations shown in Figure~\ref{fig:chandra}.

\begin{figure}
\plottwo{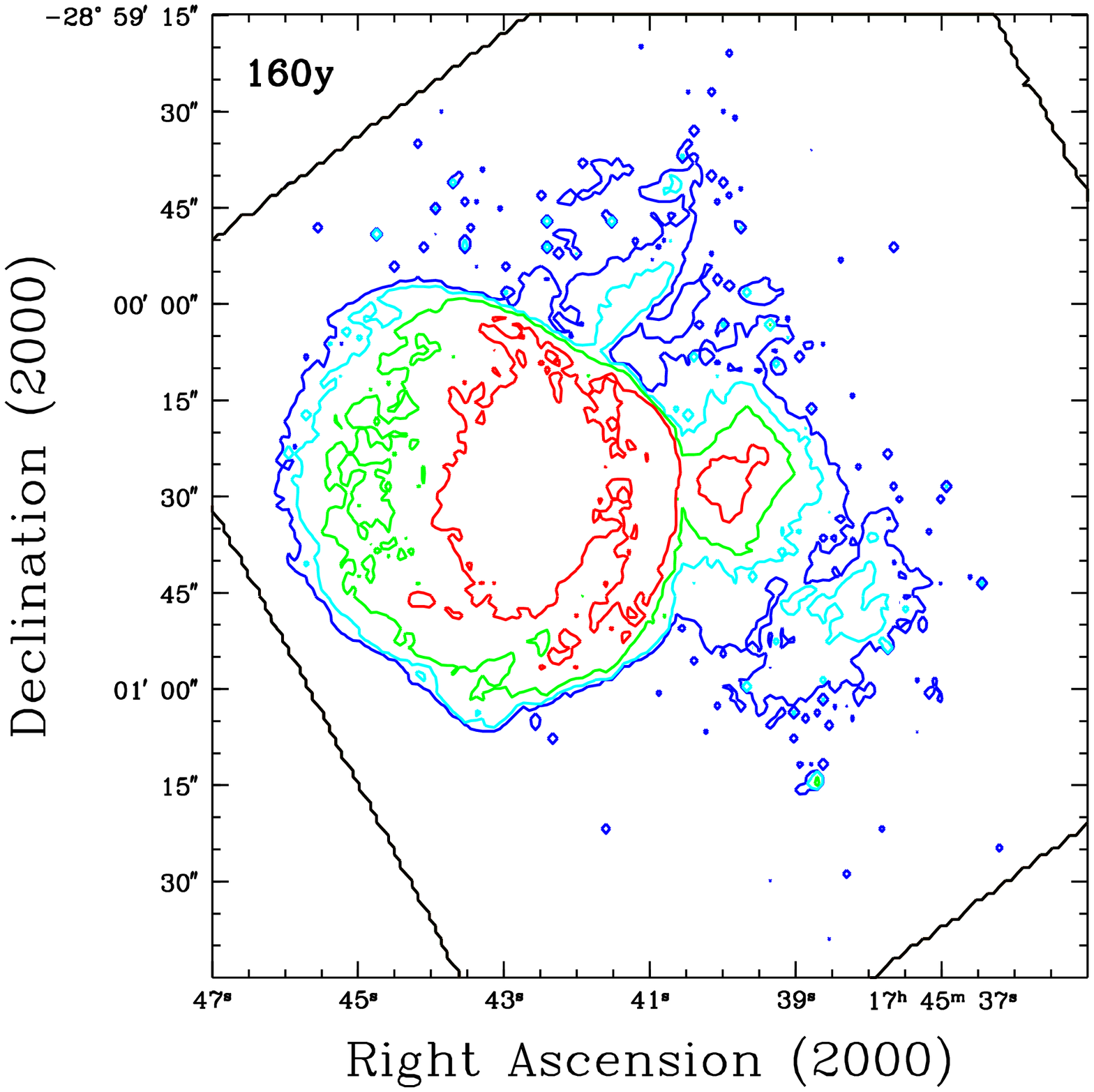}{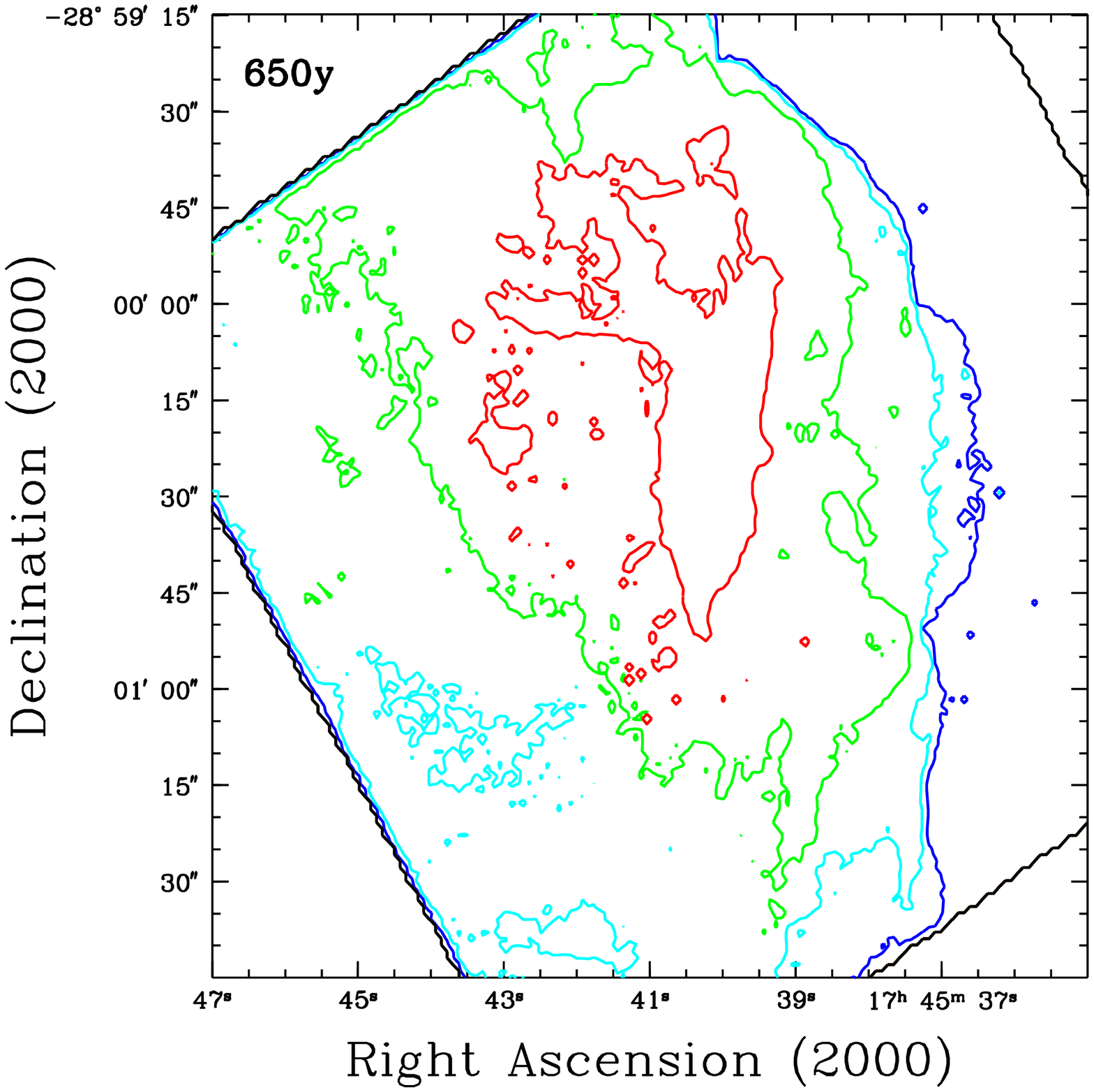}

\plottwo{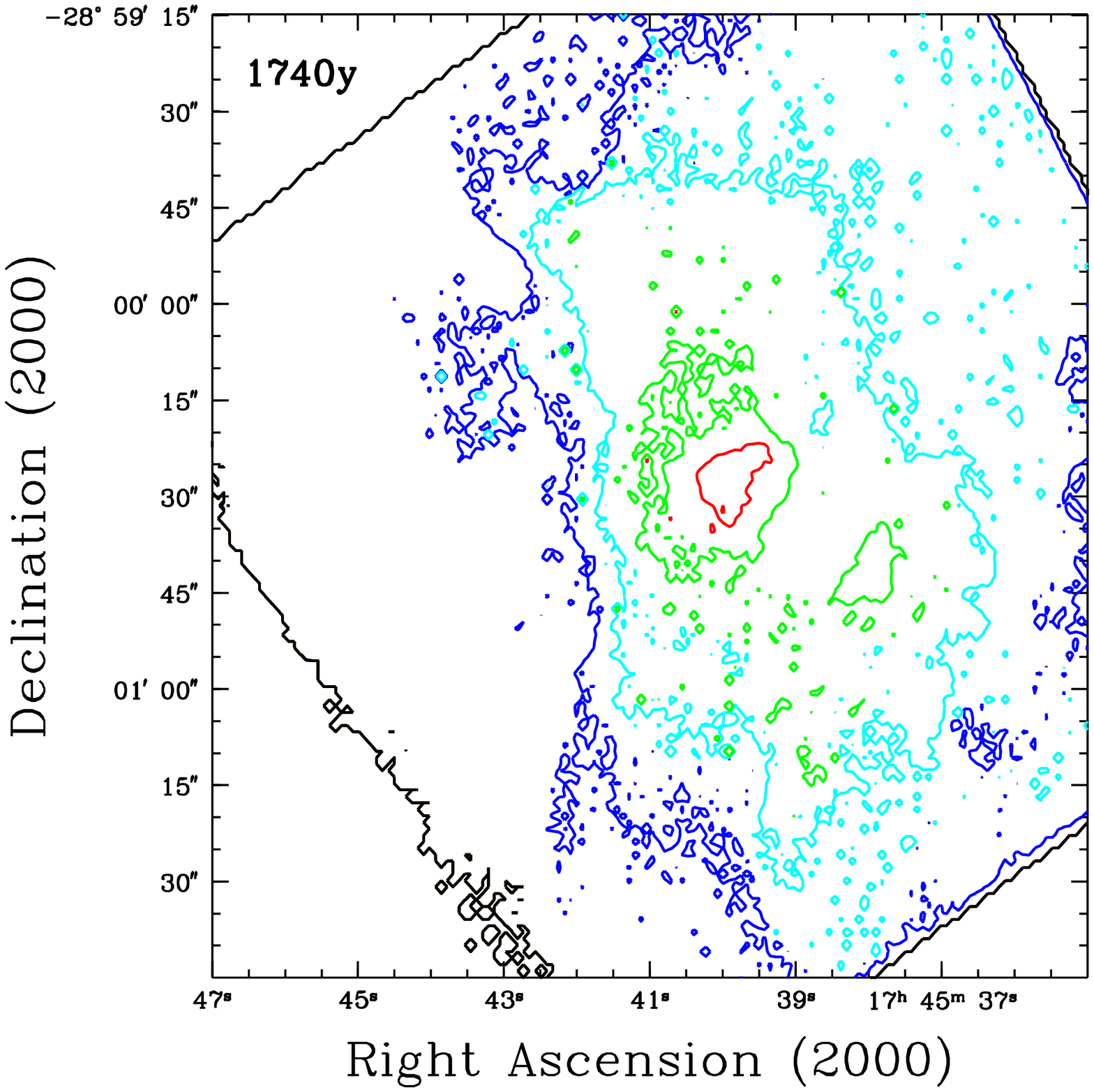}{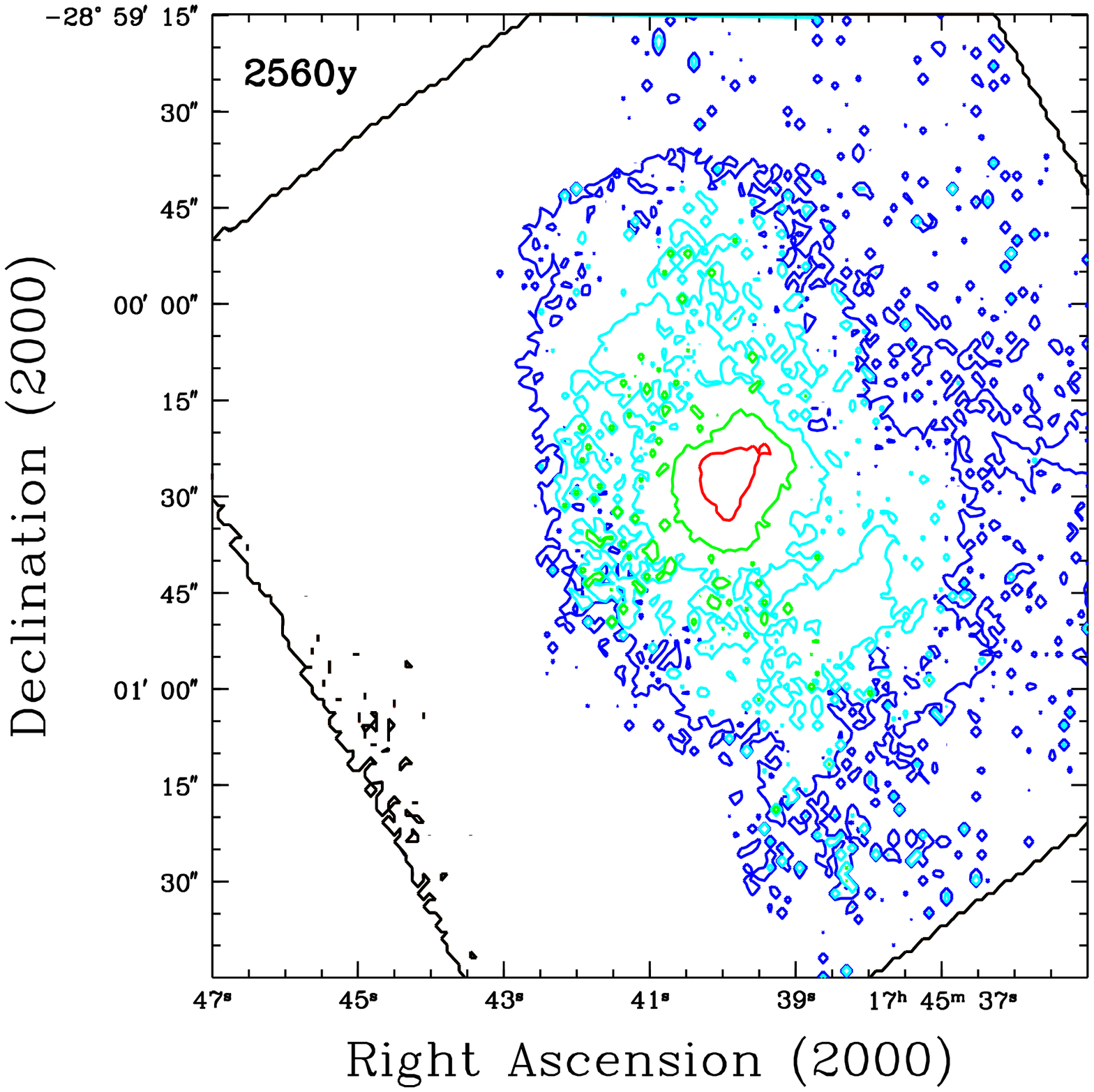}
\caption{Contours of column-integrated 2--10~keV X-ray luminosity per
  $1\arcsec.2\times1\arcsec.0$ bin from our simulation, shown at four
  different snapshots in time (160, 650, 1740, and 2560 years after
  the supernova explosion).  In order from blue to cyan to green to
  red, the luminosity densities indicated by the contours are $1.5
  \times 10^{29}$, $10^{30}$, $10^{31}$,
  $10^{32}$~erg~s$^{-1}$~bin$^{-1}$.  At 160 years, the shock from the
  supernova ejecta hitting the dense wind-dominated region surrounding
  Sgr~A* produces an X-ray hot spot.  The X-ray hot spot moves around
  Sgr~A* as the head of the supernova shock sweeps around the central
  region (650 years).  After 1740 years, the stellar wind material
  begins to reassert itself.  The shock between it and the slower
  moving supernova ejecta forms an X-ray ridge with a 2--10 keV
  luminosity of roughly $3 \times 10^{33}$~erg~s$^{-1}$.  This is our
  best estimate for the current state of the Galactic center.  With
  time, this ridge will move out and dim.  By 2560 years after the
  launch of the explosion (800 years from now), the luminosity of the
  ridge will be a factor $\sim 3$ dimmer, and the feature will lie
  nearly $30\arcsec$ away from Sgr~A*.}
\label{fig:xray}
\end{figure}

The column integrated X-ray emission from our models is shown in
Figure~\ref{fig:xray} in a series of temporal snapshots.  For very
energetic shocks, the high compression ratio at the point of impact
can lead to significant particle acceleration and consequent
nonthermal emission, but by this time in the interaction, the dominant
$2$--$10$ keV emission mechanism is expected to be optically thin
bremsstrahlung \citep[see][for details]{Roc04,Fry05}.  The supernova
shock reaches the dense inner $0.2$--$0.3$~pc of the Galactic Center
in 160 years and sweeps around this area after 650 years.  The X-ray
emission is highest where the shock is strongest (a function of both
shock velocity and density of the impacted region).

By $1,740$ years after the explosion, the supernova shock has swept
beyond our simulation grid.  However, over 95\% of the mass in the
supernova ejecta is moving at less than half the speed of the
supernova shock front.  This slow-moving material continues to impinge
on the inner $0.2$--$0.3$~pc of the Galaxy.  It is the interaction
between the stellar winds pushing against the slow-moving supernova
ejecta that produces the X-ray ridge observed today
(Fig.~\ref{fig:chandra}).

For direct comparison between our simulation and the properties
evident in Figure~\ref{fig:chandra}, we restrict our attention to the
region bounded by the $9\arcsec$ and $15\arcsec$ arcs in this image;
this appears to be the scene of dominant interaction at the present
time.  The observed flux is accurately modeled with two components at
different temperatures: a $5.6$~keV component with a 2--10~keV flux of
$3.92 \times 10^{-13}$~erg~cm$^{-2}$~s$^{-1}$ and a $1$~keV component
with a 2--10~keV flux of $5.8 \times 10^{-13}$~erg~cm$^{-2}$~s$^{-1}$;
these correspond to 2--10~keV luminosities of $3.0 \times
10^{33}$~erg~s$^{-1}$ and $4.4 \times 10^{33}$~erg~s$^{-1}$,
respectively, at a distance of 8~kpc from the Galactic center.

In our simulation, both the location of, and the X-ray intensity from,
this region vary with time, so we have essentially two important
constraints on the comparison between theory and observation.  We
suspect that the 1~keV component arises from either foreground or
background emission; this is supported by the fact that the
luminosity-weighted temperature in this region of our simulation is
$4.9$~keV after $1,740$~years.  From our simulation, we find that the
column-integrated 2--10~keV luminosity from the interaction within the
swath highlighted in Figure~\ref{fig:chandra} is $3.6 \times
10^{33}$~erg~s$^{-1}$ after $1,619$~years and $2.7 \times
10^{33}$~erg~s$^{-1}$ after $1,740$~years.  These values are within
20\% of the observed luminosity of the $5.6$~keV component, and the
good match between theory and observation for both the morphology and
the X-ray radiance of the ridge therefore provides us with compelling
evidence that $\sim 1,700$ years must be a reasonable estimate for the
remnant's age.

As the velocity of the impinging supernova ejecta decreases, the X-ray
ridge moves out and dims.  By $2,560$ years, the ridge will have moved
out $30\arcsec$ and its flux should then be 3 times lower than its
(current) $1,740$-year value.

\section{Conclusion}

Our simulation of the passage of Sgr~A East across the Galactic center
has produced several new insights into the structure of the
environment within $\sim 3$~pc of Sgr~A*.  The front of the SNR flowed
around the Galactic center $\sim 1,100$ years ago (650 years after the
supernova explosion).  The shock front pushed back the combined winds
from the Wolf-Rayet and O/B stars, but did not get within $\sim 0.2$
pc of the accreting supermassive black hole.  The collision between
the supernova ejecta and the central winds produces a ridge of
X-ray-emitting gas $\sim 9\arcsec$--$15\arcsec$ to the NE of the black
hole.  Comparing our simulations with the observations allows us to
estimate the age of the supernova remnant to be $\sim 1,700$ years.
We predict that this X-ray ridge will move out and dim with time.  In
$\sim 800$ years, it should be roughly twice its current distance from
Sgr~A*, and a factor $\sim 3$ dimmer.

The supernova did not significantly alter the accretion rate
onto the black hole.  We discuss this in greater detail in an upcoming
paper \citep{Fry05}.  Our simulations demonstrate how rich in
information are the X-ray observations of the Galactic Center.
Although this region contains several complex flows that are difficult
to simulate completely, the extensive body of data contains many
features that we can employ to tightly constrain the calculations.  As
we continue to refine our models, we expect to uncover additional new
features regarding the environment and physical processes at the
Galactic center.

This work has barely scratched the surface of what we can learn about
the Galactic center by carefully studying the gas dynamics in this
region.  In \citet{Fry05}, we will discuss in more detail the
hydrodynamic evolution of the shock and calculate the emission as the
shock hits the large molecular cloud to form Sgr~A East.  We will also
report in detail the metallicity gradients expected within Sgr~A East
as a function of time and distance from Sgr~A*, providing yet another
observational signature that may be used to better constrain this
remnant's age.

{\bf Acknowledgments} This work was funded in part under the auspices
of the U.S.\ Dept.\ of Energy, and supported by its contract
W-7405-ENG-36 to Los Alamos National Laboratory, by a DOE SciDAC grant
DE-FC02-01ER41176.  At The University of Arizona, this research was
supported by NSF grant AST-0402502, and has made use of NASA's
Astrophysics Data System Abstract Service.  F. M. is grateful to the
University of Melbourne for its support (through a Sir Thomas Lyle
Fellowship and a Miegunyah Fellowship).  The simulations were
conducted on the Space Simulator at Los Alamos National Laboratory.

{}

\newpage

\clearpage

\clearpage

\clearpage

\end{document}